\documentclass[conference]{IEEEtran}
\IEEEoverridecommandlockouts
\usepackage{cite}
\usepackage{amsmath,amssymb,amsfonts}
\usepackage{algorithmic}
\usepackage{graphicx}
\usepackage{textcomp}
\usepackage{xcolor}

\def\BibTeX{{\rm B\kern-.05em{\sc i\kern-.025em b}\kern-.08em
    T\kern-.1667em\lower.7ex\hbox{E}\kern-.125emX}}
\begin{document}

\title{LADRI: \textbf{L}e\textbf{A}rning-based \textbf{D}ynamic \textbf{R}isk \textbf{I}ndicator in Automated Driving System\\
}

\author{\IEEEauthorblockN{Anil Ranjitbhai Patel}
\IEEEauthorblockA{\textit{RPTU Kaiserslautern-Landau, Germany} \\
apatel@rptu.de}
\and
\IEEEauthorblockN{Peter Liggesmeyer}
\IEEEauthorblockA{\textit{RPTU Kaiserslautern-Landau, Germany} \\
peter.liggesmeyer@rptu.de}
}

\maketitle

\begin{abstract}
As the horizon of intelligent transportation expands with the evolution of Automated Driving Systems (ADS), ensuring paramount safety becomes more imperative than ever. Traditional risk assessment methodologies, primarily crafted for human-driven vehicles, grapple to adequately adapt to the multifaceted, evolving environments of ADS. This paper introduces a framework for real-time Dynamic Risk Assessment (DRA) in ADS, harnessing the potency of Artificial Neural Networks (ANNs).

Our proposed solution transcends these limitations, drawing upon ANNs, a cornerstone of deep learning, to meticulously analyze and categorize risk dimensions using real-time On-board Sensor (OBS) data. This learning-centric approach not only elevates the ADS's situational awareness but also enriches its understanding of immediate operational contexts. By dissecting OBS data, the system is empowered to pinpoint its current risk profile, thereby enhancing safety prospects for onboard passengers and the broader traffic ecosystem.

Through this framework, we chart a direction in risk assessment, bridging the conventional voids and enhancing the proficiency of ADS. 
By utilizing ANNs, our methodology offers a perspective, allowing ADS to adeptly navigate and react to potential risk factors, ensuring safer and more informed autonomous journeys.
\end{abstract}


\section{Introduction}
ADS, operating in dynamic environments with minimal human intervention, are necessitated to maintain constant safety measures. 
Despite traditional automotive safety frameworks like ISO 26262 using Automotive Safety Integrity Levels (ASIL) and Hazard Assessment and Risk Analysis (HARA), a comprehensive safety strategy for ADS remains elusive due to these methods' static nature and worst-case assumptions. 
IEC 31010 \cite{b2}, though offering numerous risk assessment techniques, lacks suitable methodologies for real-time risk analysis. 
ADS, frequently encountering uncertain and unfamiliar scenarios, necessitate a LeArning-based Dynamic Risk Indicator (LADRI) model that incorporates ``Risk Knowledge" to adapt and respond to real-time situations \cite{b3,b4}.
This has instigated research in fields like Runtime Certiﬁcation, Dynamic Safety Management, Runtime Monitoring, Safety Supervisor, Dynamic Safety Cases, and Conditional Safety Certiﬁcates, which hold the potential for real-time risk assessment \cite{b5}. 

The contributions of the paper lie primarily in the development of a LADRI model, which offers a more contextual, comprehensive, and dynamic approach to risk assessment in ADS. 
While conventional risk assessment methods that employ dynamical motion models for traffic participants and the ego vehicle certainly have their strengths, the LADRI model offers several unique advantages.
\begin{enumerate}
    \item \textbf{Dynamic Risk Assessment:} LADRI leverages machine learning algorithms to process a broad array of data from OBS in real-time, enabling it to respond to rapidly changing traffic situations and environmental conditions. This DRA provides more timely and accurate predictions than static, rule-based models.
    \item \textbf{Learning Capability:} As a machine learning model, LADRI is capable of learning and improving over time. With each new piece of data or experience, the model can refine its risk prediction capabilities, making it progressively more accurate and reliable.
    \item \textbf{Continuous Improvement:} LADRI’s ability to learn from past experiences and improve its risk prediction capabilities over time ensures that it stays relevant and effective as ADS technology and environmental variables evolve.
    \item \textbf{Generalization Across Scenarios:} Given its learning-based nature, LADRI is not confined to predefined rules or parameters. As a result, it can generalize its learning across different scenarios, making it adaptable and versatile in diverse real-world situations.
\end{enumerate}

\begin{figure*}[htbp]
\centering
\includegraphics[width=0.68\textwidth]{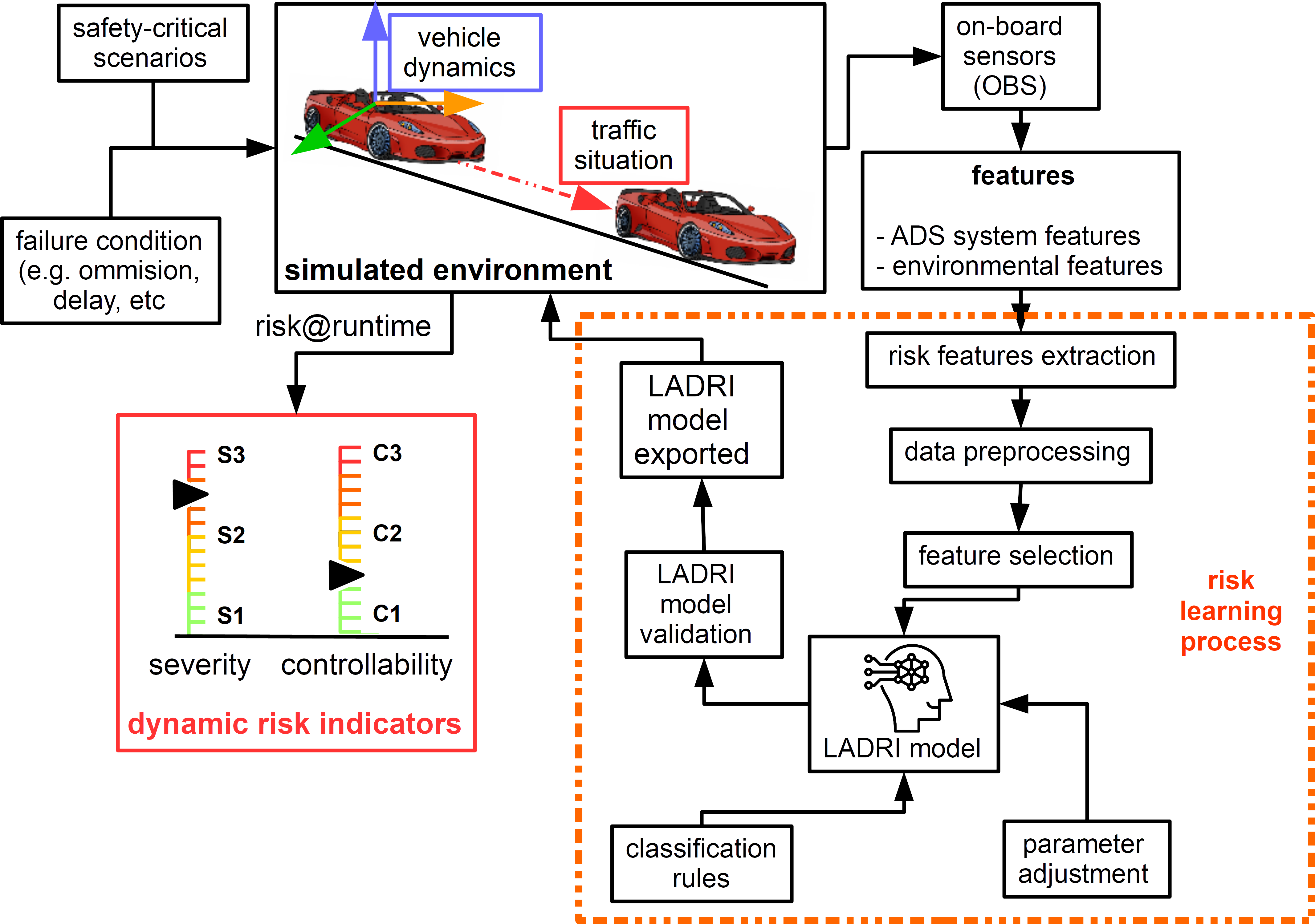}
\caption{Learning-based Dynamic Risk Indicator Framework} \label{fig1} 
\end{figure*}

\section{Related work}
Ensuring the safety of ADS when operating on public roads remains a primary concern, underscoring the significance of a comprehensive DRA. One avenue that has been explored to complement and elevate traditional risk assessment methodologies is the integration of machine learning techniques. Hegde et al.'s work is an example, where they delved into the potential of machine learning to refine risk assessment by infusing it with data-driven insights \cite{b6}. However, current models based on CNN overlook crirical driving dynamics, focusing primarily only on distance and velocity \cite{b7}.

To overcome these constraints, the Safety Monitoring Framework (SMOF) provides organized frameworks for online safety regulations  \cite{b8}. However, it encounters its own hurdles, primarily in the effective organization and management of customized, complex rules designed for autonomous systems. Similarly, the Structured Approach for HARA (SAHARA) \cite{b8a} aspires to standardize operational contexts and harmonize them with computer-assisted HARA to enhance reliability and performance. Nevertheless, it falls short in formalizing aspects of controllability and severity.

Given the obstacles, the SINADRA framework provides a structured method for real-time risk surveillance, drawing on tactical environmental insights and forward-thinking risk management tactics \cite{b9}. This framework aligns with the LADRI model's goals, especially when tackling the most severe risk assessment assumptions. SINADRA primarily utilizes Bayesian network inference for risk assessment, while LADRI employs machine learning to predict risks, notably using data from OBS in complex driving scenarios. Combined, these advancements signify a move towards flexible, data-driven risk evaluation processes for ADS safety.

\section{Methodology}
The LADRI model presents an approach to DRA within ADS. Its distinguishing factor from the traditional HARA method is its streamlined efficiency. While the HARA process relies heavily on wide-ranging professional inputs and exhaustive expert consultations, LADRI has been designed to be more agile and direct.
At its core, LADRI leverages real-time data from OBS and combines this with advanced analytical capabilities to provide a dynamic risk indicator, even in the most fluid of driving environments. 
This is underpinned by the use of Artificial Neural Networks (ANNs), which grants LADRI its adaptability. Depending on the specific data input and the challenges at hand, LADRI can seamlessly switch between various neural network architectures. This could range from RNNs, LSTMs, and FNNs to supervised classification algorithms like Support Vector Machine \cite{b10}, decision trees \cite{b11}, or random forests.

This versatility allows LADRI to be effectively tailored to the distinct demands of diverse situations, ensuring optimal performance. Moreover, while we acknowledge concerns over potential discrepancies in sensor readings or other ADS malfunctions, the strength of LADRI lies in its comprehensive data processing capability. It's adept at sifting through extensive data, pinpointing patterns, and highlighting anomalies, whether they stem from a faulty sensor or other potential risks.

By utilizing data from OBS in real-time, LADRI identifies pertinent hazardous events and quantifies the severity of potential harm (S) as well as controllability (C). This method not only validates the adherence to predetermined safety goals during adverse events but also provides a clearer picture of the rigor required in the development of safety mechanisms for specific conditions.

The learning-process involves data preprocessing, risk feature extraction, parameter adjustments, and defined classification rules for fine-tuning the model as shown in Fig. \ref{fig1} (adapted from \cite{b10}). The validation of the LADRI model is comprehensive, employing techniques such as dataset splitting, cross-validation, performance metrics evaluation, and comparison with baseline models. Through rigorous validation, we ensure the reliability and effectiveness of the LADRI model in assessing dynamic risk.

In practical application, such as in an Adaptive Cruise Control (ACC) use case, the LADRI model utilizes data from a wide range of OBS, such as radar, LiDAR, ultrasonic sensors, wheel speed sensors, engine speed sensor, Throttle-Pedal position sensor, and Brake-Pedal position sensor. This multi-sensor data serves as an input for LADRI model, and the model extracts relevant risk features, such as relative distance and speed, traffic density, vehicle dynamics, road type and conditions, lane departure indicators, vehicle control inputs, and time-to-collision. Upon validation, the model is integrated into a high-fidelity driving simulator for real-time risk indicators. It is tested across diverse safety-critical scenarios (e.g. unintended acceleration or unintended braking), providing valuable insights into its performance and ability to predict dynamic risk accurately and timely in changing scenarios.

This research aims to contribute to the existing body of knowledge by offering a learning-based DRA tool that can adapt to changing data and driving conditions, thereby providing a more accurate dynamic risk indicator in ADS.

\begin{figure}[htbp]
\centering
\includegraphics[width=0.36\textwidth]{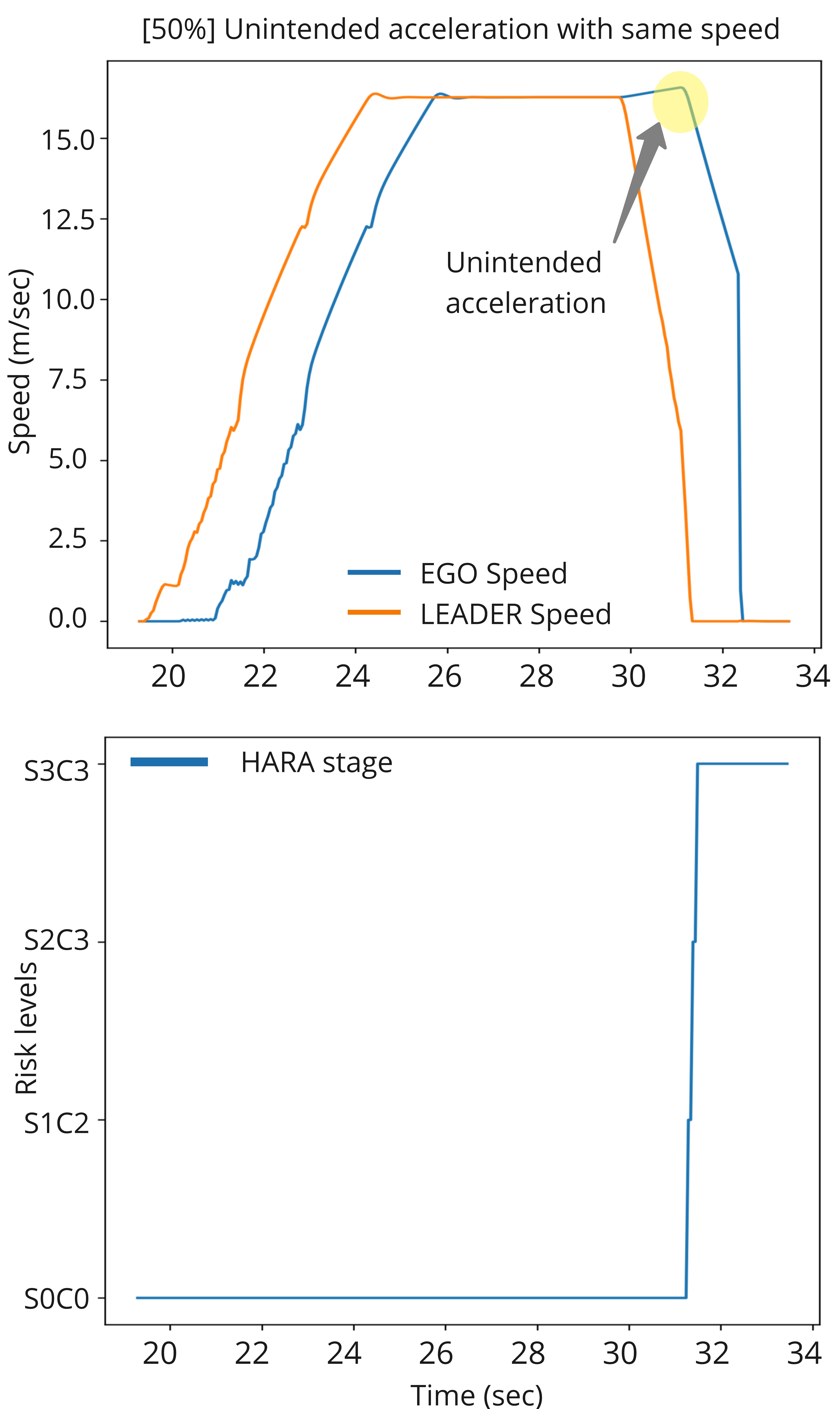}
\caption{Unintended Acceleration with Varying Throttle Value} \label{fig3}
\end{figure}

\section{Results and Discussion}
In Fig. \ref{fig3}, we analyze a specific scenario where the ego vehicle follows a lead vehicle that maintains a steady speed of 60 km/h. One of the most pronounced observations from this scenario is the effect of unintended acceleration on the dynamics between the two vehicles. Notably, when the unintended acceleration spikes by 50\%, the transition to a critical state is markedly faster. This swift shift towards the critical state can be attributed to the significant variation in both speed and distance between the two vehicles, as evidenced by the figure.

Several important arguments can be derived from this analysis:

1. Sensitivity to Unintended Acceleration: Even a moderate increase in unintended acceleration (in this case, 50\%) can dramatically alter the dynamic interplay between vehicles, emphasizing the sensitivity of the system to acceleration deviations.

2. Safety Implications at Higher Speeds: The scenario underscores that unintended accelerations at higher speeds (like 60 km/h) can exponentially amplify safety risks. The combined effect of high speed and abrupt acceleration can reduce the reaction time available for the following vehicle, making collision avoidance more challenging.

3. Importance of Distance Maintenance: A rapid progression to the critical state, as induced by unintended acceleration, emphasizes the importance of maintaining safe following distances. Sudden changes in speed can erode these distances quickly, leaving little margin for error.

4. Predictive Safety Measures: The findings highlight the need for ADS to incorporate predictive safety measures that can preemptively identify and rectify such unintended accelerations, thereby preserving safe vehicular dynamics.

In conclusion, the results from Fig. \ref{fig3} serve as a strong reminder of the intricate dynamics at play in vehicular scenarios and the paramount importance of ensuring that ADS systems can effectively navigate and mitigate these complexities.

\begin{figure}[htbp]
\centering
\includegraphics[width=0.36\textwidth]{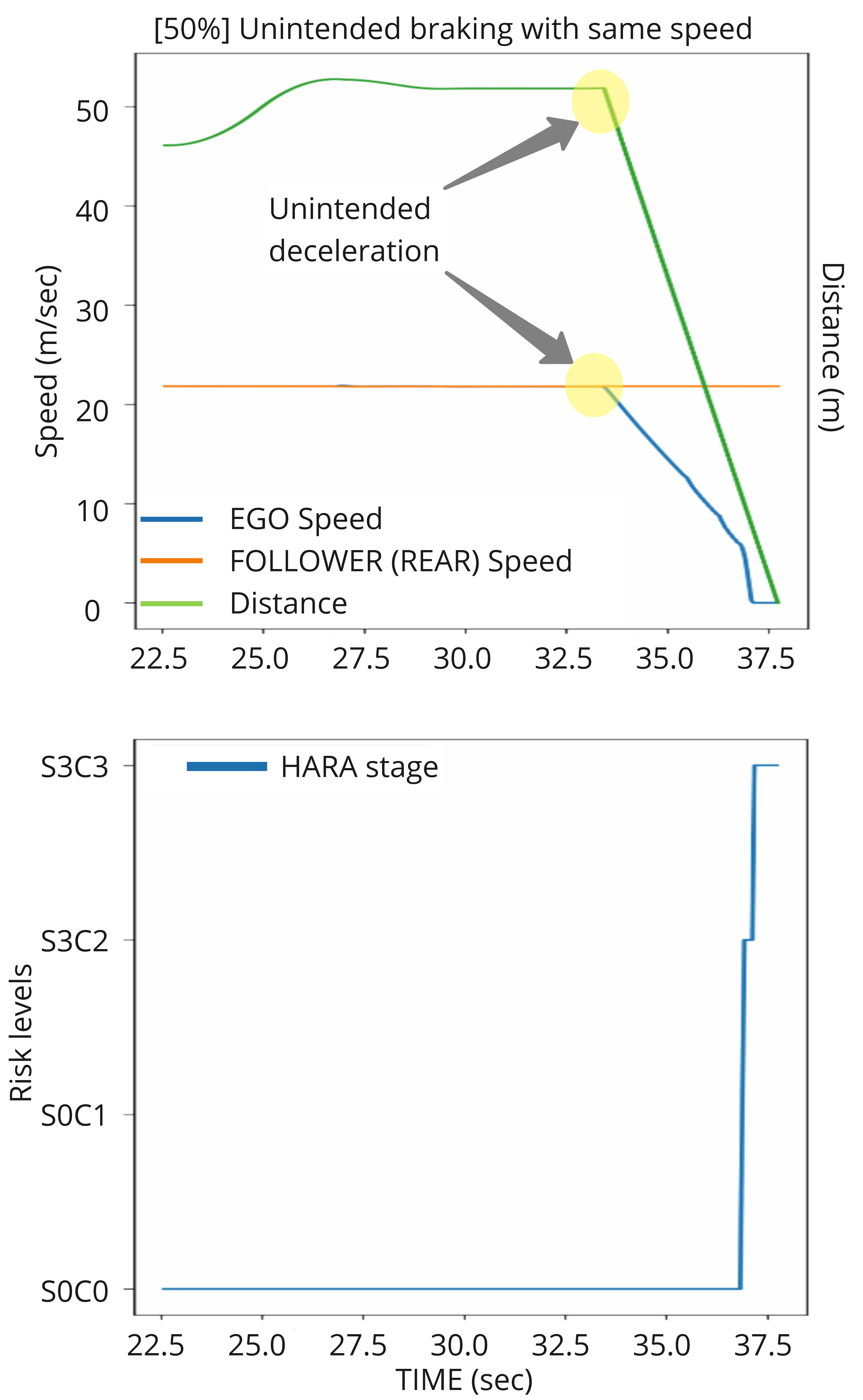}
\caption{Unintended Braking with Varying Brake Value} \label{fig4}
\end{figure}

In Fig. \ref{fig4}, we observe a highway scenario featuring an ego vehicle (represented in Blue) and a subsequent follower vehicle (depicted in Orange). Within this context, when the follower vehicle matched the speed of the ego vehicle, the HARA stage experienced a noticeable surge. Intriguingly, a sudden braking incident was initiated by the system in the absence of any discernible obstruction, prompting an aggressive deceleration. This spontaneous system response not only escalated the HARA stage but also underscored the profound influence of the follower vehicle's speed on risk assessment parameters.

From this illustrative scenario, we can extrapolate several salient points:

1. Synchronization and Implications: The synchronization of speeds between the ego and the follower vehicle, as seen in the scenario, can lead to heightened sensitivity in the HARA metrics. The intricacies of vehicular dynamics come to the fore, with speed matching potentially elevating risk levels.

2. Unanticipated Braking Dynamics: The unexpected braking action, devoid of any apparent trigger, is a testament to potential anomalies or over-reactiveness in the system. Such unpredictable behaviors can introduce substantial risks, especially on highways where vehicles typically maintain higher speeds.

3. Impact of Follower Vehicle's Speed: The scenario underscores the pivotal role played by the follower vehicle's speed in risk assessment. Speed congruence or disparity between vehicles can significantly modulate the HARA stages, revealing the intricate relationship between vehicular speeds and resultant risk profiles.

4. Advocacy for Robust System Calibration: The insights gleaned emphasize the importance of a finely-tuned and calibrated system. ADS should possess the capability to discern genuine obstructions from false positives, ensuring appropriate and proportionate responses to actual threats.

In summation, Fig. \ref{fig4} serves as a compelling exposition of the nuanced dynamics between vehicles on highways. It underlines the imperative for advanced driving systems to be both perceptive and judicious in their responses, safeguarding against potential hazards while maintaining fluid vehicular dynamics.

Findings indicated that the LADRI model accurately captured and responded to these changes, validating its effectiveness in predicting dynamic risk. However, it was acknowledged that further research is needed to address the limitations of ANN models, specifically their lack of explanation mechanisms and white-box style prediction capabilities.

In conclusion, the results suggest that the LADRI model, utilizing OBS data, can contribute to improving the safety of ADS by raising the level of automation from fail-safe to fail-operational. Future enhancements will focus on improving the model's explainability to ensure its applicability in real-world scenarios. 

\section{Limitation of LADRI}
While the LADRI model offers DRA in autonomous driving scenarios, it faces challenges in ensuring a comprehensive understanding of both functional and non-functional requirements in unpredictable environments. The integration of varied sensors and subsystems, crucial for accurate risk modeling, introduces complexity, especially as component interactions grow. Decomposing the system, while aiding in manageability, may inadvertently create dependencies and potential failure points. Moreover, crafting models that encapsulate the vast spectrum of driving conditions, regulatory compliance, and real-time interactions amplifies the computational demands, potentially impinging on the framework's real-time efficacy and efficiency.

\section*{Acknowledgment}
This paper discusses research conducted within the framework of a broader project sponsored by the Performance Center Simulation and Software-based Innovation (Leistungszentrum Simulations- und Software-basierte Innovation) based in Kaiserslautern, Germany. I would also like to express my gratitude for the unwavering support provided by my student, Sanjay Gorasiya, and my colleague, Nikita Bhardwaj-Haupt.



\begin{thebibliography}{00}
\bibitem{b2} ISO/IEC-31010:2019 Risk management - Risk assessment techniques
\bibitem{b3} T. Aven and B. S. Krohn, "A new perspective on how to understand, assess and manage risk and the unforeseen," Reliability Engineering and System Safety, vol. 121, pp. 1-10, 2014.
\bibitem{b4} N. Paltrinieri, L. Comfort, and G. Reniers, "Learning about risk: Machine learning for risk assessment," Safety Science, vol. 118, pp. 475-486, 2019.
\bibitem{b5} A. R. Patel, N. B. Haupt, and P. Liggesmeyer, "A Conceptual Framework of Dynamic Risk Management for Autonomous Vehicles," in New Trends in Intelligent Software Methodologies, Tools and Techniques. IOS Press, 2022, pp. 475-486.
\bibitem{b6} J. Hegde and B. Rokseth, "Applications of machine learning methods for engineering risk assessment–A review," Safety Science, vol. 122, pp. 104492, 2020.
\bibitem{b7} P. Feth, M. N. Akram, R. Schuster, and O. Wasenmüller, "Dynamic risk assessment for vehicles of higher automation levels by deep learning," in SAFECOMP 2018 Workshops, Västerås, Sweden, September 18, 2018.
\bibitem{b8} M. Machin, J. Guiochet, H. Waeselynck, J. P. Blanquart, M. Roy, and L. Masson, “SMOF: A safety monitoring framework for autonomous systems,” IEEE Trans. Syst., Man, Cybern.: Syst., vol. 48, no. 5, pp. 702–715, 2016.
\bibitem{b8a} S. Kemmann, "SAHARA-a structured approach for hazard analysis and risk assessments," Doctoral dissertation, Technische Universität Kaiserslautern, 2015.
\bibitem{b9} J. Reich and M. Trapp, "SINADRA: Towards a framework for assurable situation-aware dynamic risk assessment of autonomous vehicles," in 2020 16th European Dependable Computing Conference (EDCC), September 2020, pp. 47-50.
\bibitem{b10} A. R. Patel and P. Liggesmeyer, "Machine learning based dynamic risk assessment for autonomous vehicles," in 2021 International Symposium on Computer Science and Intelligent Controls (ISCSIC), Nov. 2021, pp. 73-77.
\bibitem{b11} A. R. Patel, N. B. Haupt, Liggesmeyer, and P., "Prediction of Dynamic Adaptation Technique for Autonomous Vehicles using Decision Trees," in 29th Safety-Critical Systems Symposium, 2021.
\end{thebibliography}
\end{document}